\documentclass[modern] {aastex63-altbib}

\usepackage{natbib}
\bibpunct{}{}{,}{s}{}{,}

\hypersetup{linkcolor=black,citecolor=black,filecolor=cyan,urlcolor=magenta}

\def\degr {\hbox{$^\circ$}}
\def\percc {$\mathrm{cm^{-3}}$} 

\def\Bpos{${B}_{\rm{}pos}$} 
\def\NH{${N}_{\rm{}H}$} 
\def\AV{$A_{V}$} 

\shorttitle{Magnetic Fields in Serpens South}
\shortauthors{Pillai et al.}

\begin{document}

\title{Magnetized filamentary gas flows feeding
  the young embedded cluster in Serpens South}




\correspondingauthor{Thushara G.S. Pillai}
\email{tpillai.astro@gmail.com}

\author[0000-0003-2133-4862]{Thushara G.S. Pillai}
\affiliation{Institute for Astrophysical Research, Boston University, 725 Commonwealth Avenue, Boston MA, 02215, USA}
\affiliation{Max-Planck-Institut f\"ur Radioastronomie, Auf dem H\"ugel
69, D-53121 Bonn, Germany}

\author[0000-0002-9947-4956]{Dan P. Clemens}
\affiliation{Institute for Astrophysical Research, Boston University, 725 Commonwealth Avenue, Boston, MA 02215, USA}

\author{Stefan Reissl}
\affiliation{Universit\"at Heidelberg, Zentrum f\"ur Astronomie,
  Institut f\"ur Theoretische Astrophysik, Albert-Ueberle-Str. 2,
  69120 Heidelberg, Germany}

\author[0000-0002-2885-1806]{Philip C. Myers}
\affiliation{Harvard-Smithsonian Center for Astrophysics, 60
Garden Street, Cambridge, MA 02138 USA}

\author[0000-0002-5094-6393]{Jens Kauffmann}
\affiliation{Haystack Observatory, Massachusetts Institute of
  Technology, Westford, MA 01886, USA}

\author{Enrique Lopez-Rodriguez}
\affiliation{SOFIA Science Center, NASA Ames Research Center, Moffett
   Field, CA 94035, USA}

\author{F. O. Alves}
\affiliation{Max-Planck-Institut für extraterrestrische Physik,
  Giessenbachstr. 1, 85748 Garching, Germany}

\author{G. A. P. Franco}
\affiliation{Departamento de Física–ICEx–UFMG, Caixa Postal 702, 30.123-970 Belo Horizonte, Brazil}

\author[0000-0001-9656-7682]{Jonathan Henshaw}
\affiliation{Max-Planck-Institute for Astronomy, Koenigstuhl 17, 69117 Heidelberg, Germany}

\author[0000-0001-6459-0669]{Karl M. Menten}
\affiliation{Max-Planck-Institut fuer Radioastronomie, Auf dem H\"ugel
69, D-53121 Bonn, Germany}

\author{Fumitaka Nakamura}
\affiliation{National Astronomical Observatory of Japan, 2-21-1 Osawa, Mitaka, Tokyo 181-8588, Japan}

\author[0000-0002-0368-9160]{Daniel Seifried}
\affiliation{Universit\"at zu K\"oln, I. Physikalisches Institut, Z\"ulpicher Str. 77, 50937 K\"oln, Germany}

\author{Koji Sugitani}
\affiliation{Graduate School of Natural Sciences, Nagoya City University,Mizuho-ku, Nagoya, Aichi 467-8501, Japan}

\author[0000-0002-5135-8657]{Helmut Wiesemeyer}
\affiliation{Max-Planck-Institut fuer Radioastronomie, Auf dem H\"ugel
69, D-53121 Bonn, Germany}

\begin{abstract}
Observations indicate that molecular clouds are strongly
magnetized, and that magnetic fields influence the formation of
stars. A key observation supporting the conclusion that
molecular clouds are significantly magnetized is that the orientation of
their internal structure is closely related to that of the magnetic
field. At low column densities the structure aligns parallel with the
field, whereas at higher column densities, the gas structure is
typically oriented perpendicular to magnetic fields, with a transition
at visual extinctions $A_V\gtrsim{}3~\rm{}mag$. Here we use far-infrared polarimetric observations from the HAWC+ polarimeter on SOFIA to report the discovery of a further
transition in relative orientation, i.e., a return to parallel alignment at
$A_V\gtrsim{}21~\rm{}mag$ in parts of the Serpens South cloud.  This
transition appears to be caused by gas flow and indicates that magnetic
supercriticality sets in near $A_V\gtrsim{}21~\rm{}mag$, allowing
gravitational collapse and star cluster formation to occur even in the
presence of relatively strong magnetic fields.
\end{abstract}

\section*{Introduction} \label{sec:intro}

A fundamental question in star formation physics is ``What processes
create and support dense interstellar medium (ISM) filaments and regulate 
the star formation within
them?''  While magnetic fields are predicted
to play an important role in the formation of dense filamentary
structures, of dense cores, and ultimately of stars within them, the
importance of magnetic fields relative to turbulence and gravity remains
poorly constrained \citep{crutcher2012,li2014:ppvi_bfield}. 

Dust emission polarization observations at 353\,GHz by the {\it Planck} satellite have
shown ordered magnetic field structures towards the Gould Belt clouds in the solar
neighborhood \citep{Planck-Collaboration2016}.  Studies have further
quantified the importance of magnetic fields via an assessment of  the 
relative orientation between
column density (\NH) structures (filaments) and the plane of the
sky magnetic field (\Bpos).  Within a star forming complex, low
column density features are preferentially aligned parallel to the
magnetic field, while at high column density, they tend to be perpendicular to the
field. A transition between the modes is observed for an $A_V$ range of $\simeq
2.7-3.5$\,mag \citep{Planck-Collaboration2016,soler2017}. Optical and
near-infrared (NIR) dust extinction polarization observations
of lower column density regions of filaments show similar behaviour
\citep{sugitani2011,Palmeirim2013,Franco2015,santos2016b,soler2019}. 

Yet, {\it Planck's} $\sim{}10\arcmin$ beam, which corresponds to 
linear scales of $\sim{}1~{\rm{}pc}\cdot{}(d/345~{\rm{}pc})$, was 
unable to resolve details of
the magnetic field structure on dense cores size scales ($< 0.1$\,pc). The
specific role of magnetic fields in shaping how the
filaments fragment into cores and on to form
stars and clusters thus remains unknown. 

The Serpens South cloud resides in the Aquila Rift complex
\citep{gutermuth2008} at a distance of 436 pc\,\citep{ortiz-leon2018}. 
Figure~\ref{fig:overview} shows that this cloud
harbors a prominent hub-filament system (HFS), i.e. a system of filaments that
radiate from a denser hub (a zone
with protostellar formation\citep{myers2009:fils}). 
Based on its high ratio of protostellar (Class 0/I) sources relative to
pre-main sequence (Class II/III) sources,  the star cluster in the Serpens South hub is likely the youngest cluster in the local neighborhood \citep{gutermuth2008}. 
The extreme youth and proximity of this HFS makes it an ideal laboratory
for testing the role of magnetic fields in a filamentary
dark cloud in an early stage of star cluster formation.  

\clearpage

\section*{The Data} \label{sec:obs}

Serpens South was observed using the HAWC+ polarimeter on
the 2.7-m SOFIA telescope, using the far-infrared (FIR)
$E$-band ($\lambda_{C} = 214$~$\mu$m, $\Delta \lambda = 44$~$\mu$m; see Methods).  The main
data products were the fractional polarization ($P$), position angle
(PA),  and their uncertainties ($\sigma_{P}$, $\sigma_{PA} $).

For regions of the cloud with weak detections, it was necessary to smooth the data to achieve adequate signal-to-nose ratios in polarization (PSNR~$\equiv{}P^{\prime}/\sigma_P$), where, $P^{\prime}$ is the debiased fractional polarization and $\sigma_{P}$, its uncertainty (see Methods). Radiative aligned torques \citep[RATs:][]{lazarian2007} cause dust grains in dense clouds to align such that their major axes, which are sensed by FIR \textbf{emission} polarimetry, are oriented perpendicular to the magnetic field \citep{Andersson2015}. Therefore, the HAWC+ vectors discussed throughout the text have been otated by 90$\degr$ relative to the electric field orientation to denote the plane of the sky magnetic field \Bpos.  The resultant polarization vectors are shown in Figure~\ref{fig:overview}.

The NIR \textbf{extinction} polarization vectors presented in Figure~\ref{fig:overview} were adopted from literature\citep{sugitani2011}, selecting their $H$-band data. Only polarization values that satisfied the additional quality criteria discussed in Kusune et al\citep{kusune2019} were considered.  The NIR vectors directly sense \Bpos\ and so have not been rotated.

Additionally, we used a {\it Herschel} gas (H$_2$) column density map derived
by the Herschel Gould Belt Survey (HGBS\cite{andre2010:filaments}). The map was
based on {\it Herschel} 70 to 500\,$\mu$m images\cite{Konyves2015} at a
resolution of 36\,\arcsec{}.  We used
the standard conversion factor\citep{kauffmann2008} between column density
and visual extinction, 
$N_{\rm H_{2}}= 9.4\times10^{20}$~$\rm{cm}^{-2}$~
($A_{\rm V}$~{mag})$^{-1}$.

\section*{Results}
\begin{figure*}
\begin{tabular}{ll}
\includegraphics[height=0.33\textheight,angle=0]{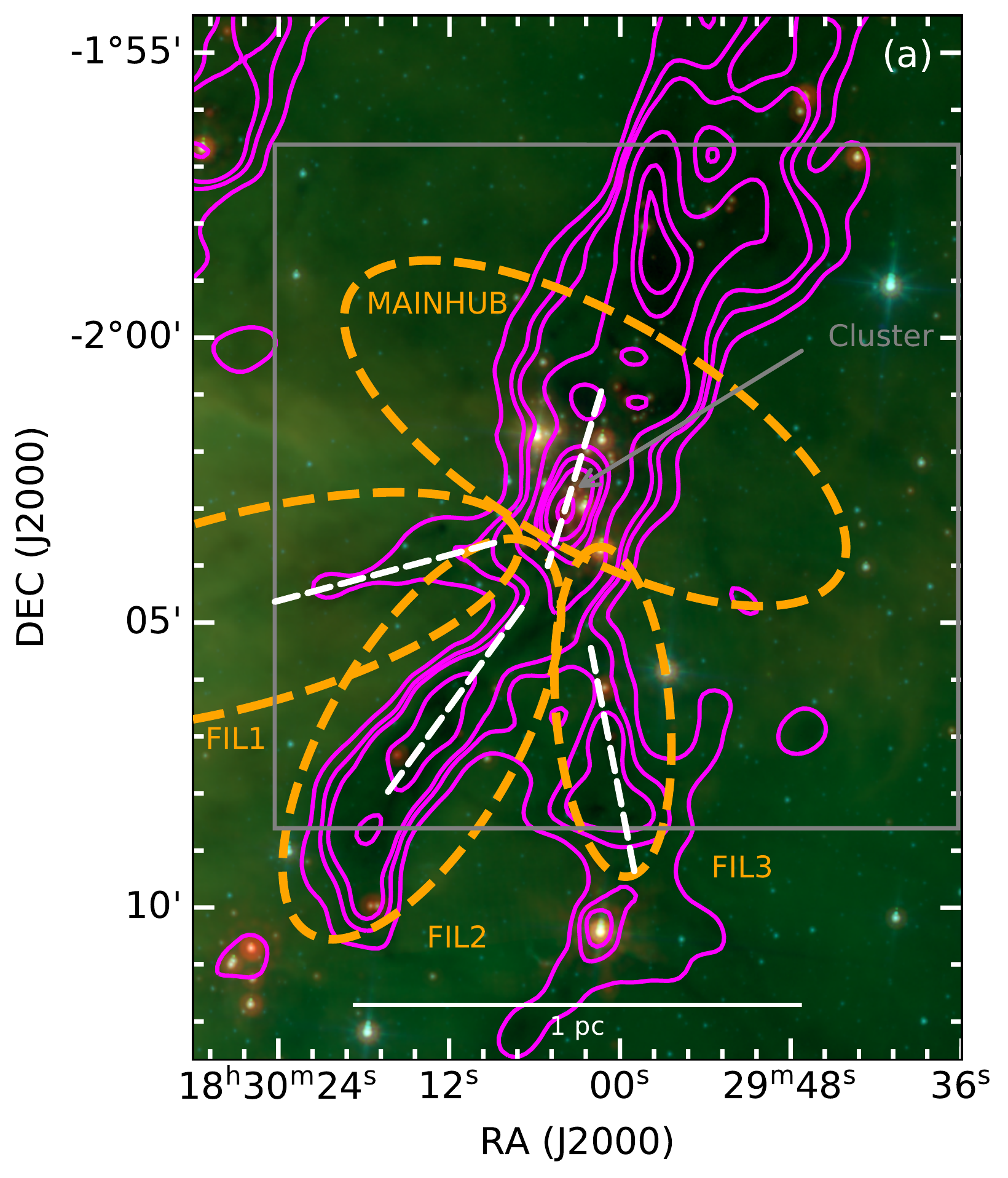}  & \includegraphics[height=0.33\textheight,angle=0]{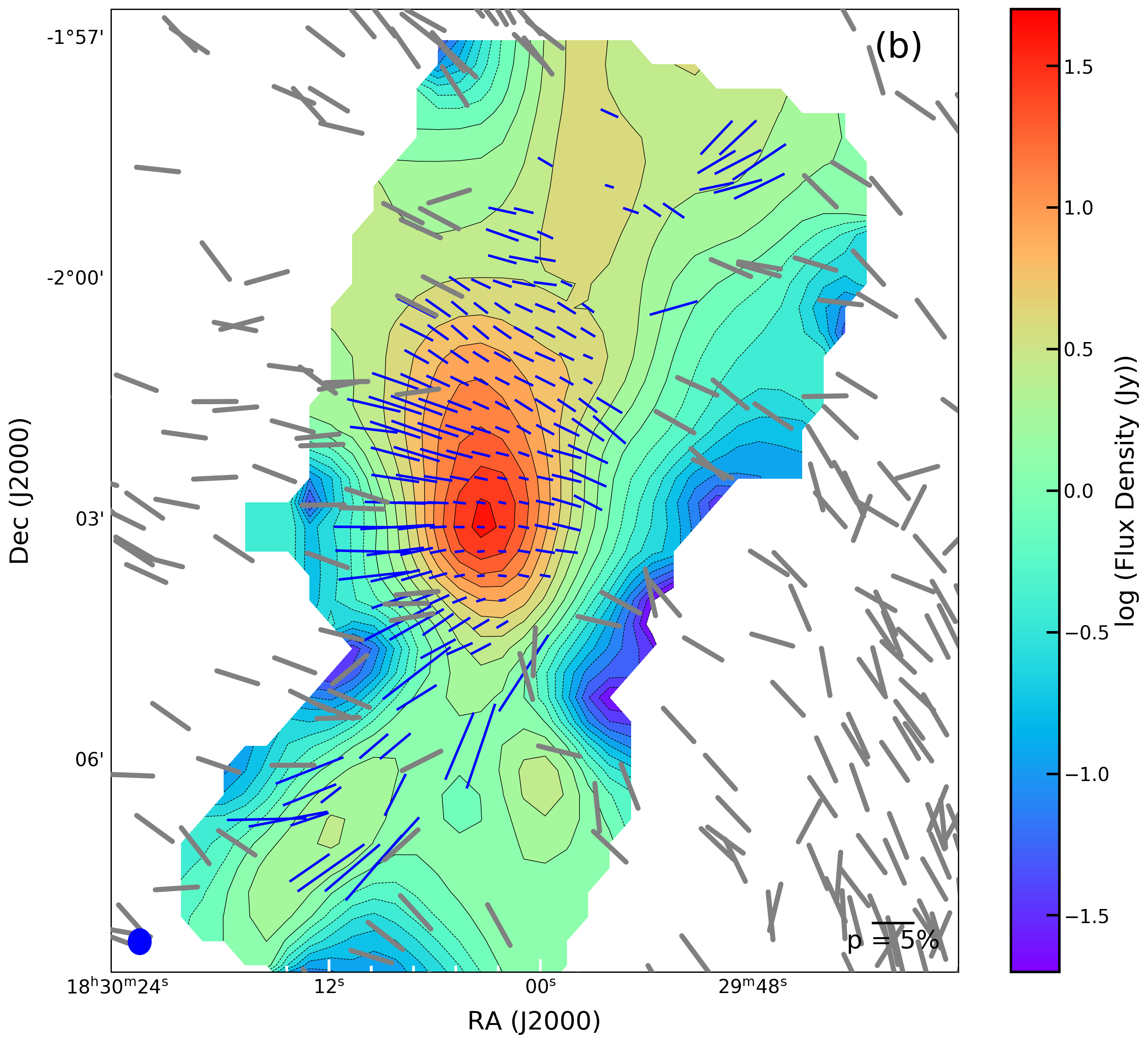}
\\
\end{tabular}
\centering
\caption{The Serpens South cloud as seen by Spitzer (a), HAWC+ intensity imaging (color scale in panel b), and polarimetry from NIR
    and HAWC+ data (vectors in b). The three--color overview map
  is generated from data acquired with the MIPS 24~$\mu$m (red), IRAC
  8~$\mu$m (blue) and IRAC 5.8~$\mu$m (green) sensors. The grey box in
  (a) shows the region mapped with HAWC+, magenta contours
  correspond to the H$_2$ column densities from {\it
    Herschel} data\cite{andre2010:filaments} at $A_V$ values of 15,
  20, 30, 45, 70, 85, 110, and 150~mag, and the dashed white lines
  show the median RHT-traced filament orientations (see text). 
    Orange ellipses delimit the regions containing MAINHUB, FIL1 and
    FIL2. White solid line of 1\,pc shows a physical length scale of 3.26 light
  years. Panel (b) presents HAWC+ 214~$\mu$m intensity (color), and polarimetry from NIR\cite{sugitani2011} (grey) and HAWC+ 214~$\mu$m data (blue, this work) at PSNR$>3$ and PSNR$>2$, respectively, tracing
  the magnetic field orientations,  corresponding to $\sigma_{PA} < 10
    (14)^{\degr}$. The blue circle at lower left shows the HAWC+ 214~$\mu$m beamsize. The
  reference percentage polarization length for HAWC+ is shown in the
  lower center, while the lengths of the NIR vectors were set to be
  identical.}
\label{fig:overview}
\end{figure*}

Figure~\ref{fig:overview} shows the Serpens South cloud,  with
NIR and HAWC+ magnetic field orientations overlaid as pseudo (headless) vectors. 
We partitioned the sky presentation of the cloud system into a hub (MAINHUB) that 
harbors the Serpens South Star Cluster and into three filaments (FIL1, FIL2,
FIL3) that appear
to connect to the hub (see Figure~\ref{fig:overview}). 
We delineated the extents of these filament regions using
ellipses. The ellipses were chosen to
maximize the number of significant polarization detections for each
region but to have minimal overlap between adjacent ellipse regions (Methods). 
The FIL3 region was removed from the remaining analysis steps as
it contained too few HAWC+ polarimetric detections. 

The large-scale magnetic field orientation follows a generally NE-SW direction that
is mostly perpendicular to the gas distribution within the MAINHUB region. 
The magnetic field orientations with respect
to the gas distributions within FIL1 and FIL2 are more complex. 
The large-scale magnetic field traced by NIR polarimetry is mostly perpendicular
to the gas filaments. But, the magnetic field orientations seen on the smaller scales 
traced by the new FIR polarimetry do not appear to follow the same patterns seen 
on the larger, NIR-traced, scales.

\begin{figure}[h]
\begin{tabular}{ll}  
\includegraphics[width=0.5\linewidth,angle=0]{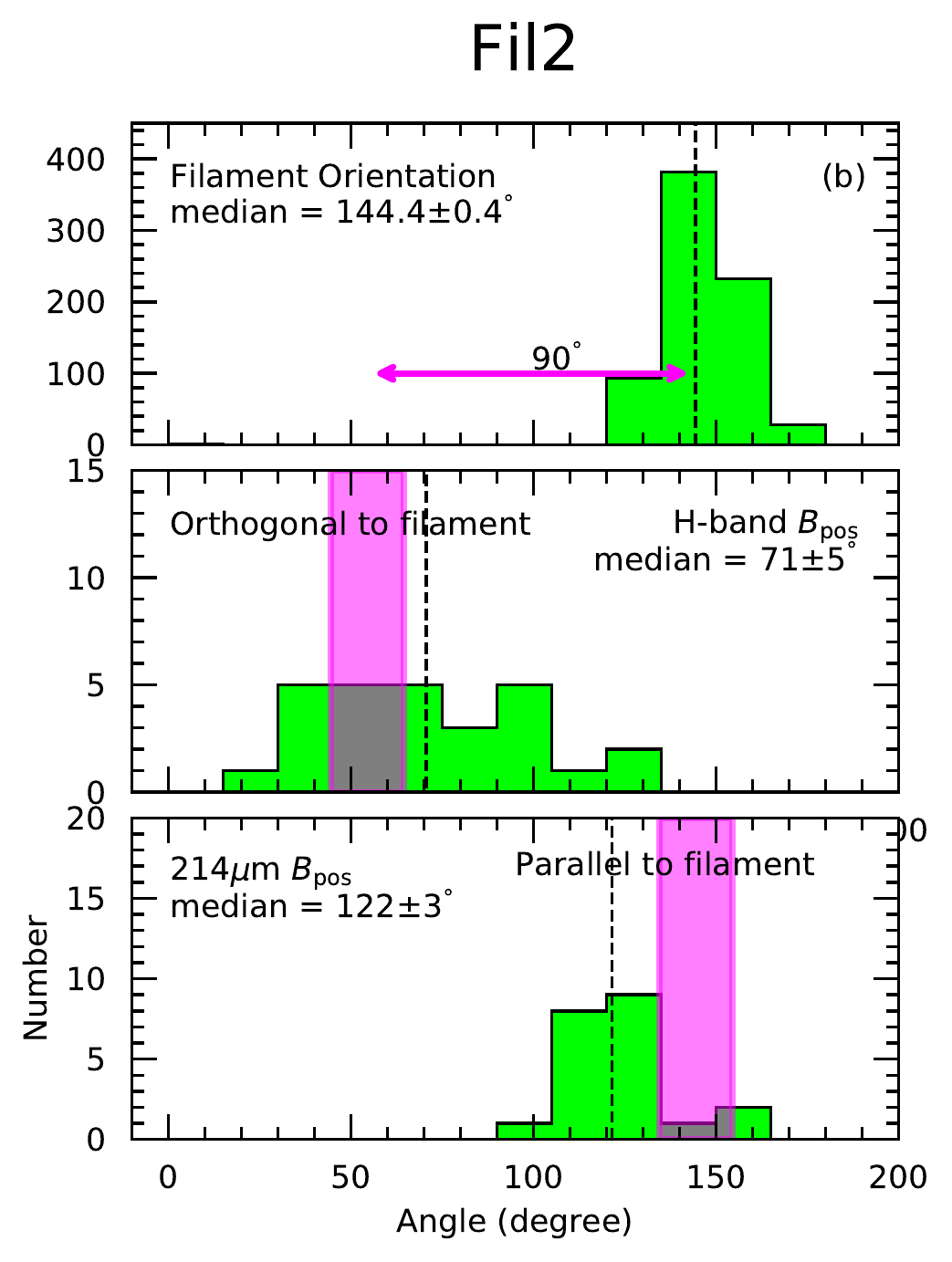} & 
\includegraphics[width=0.5\linewidth,angle=0]{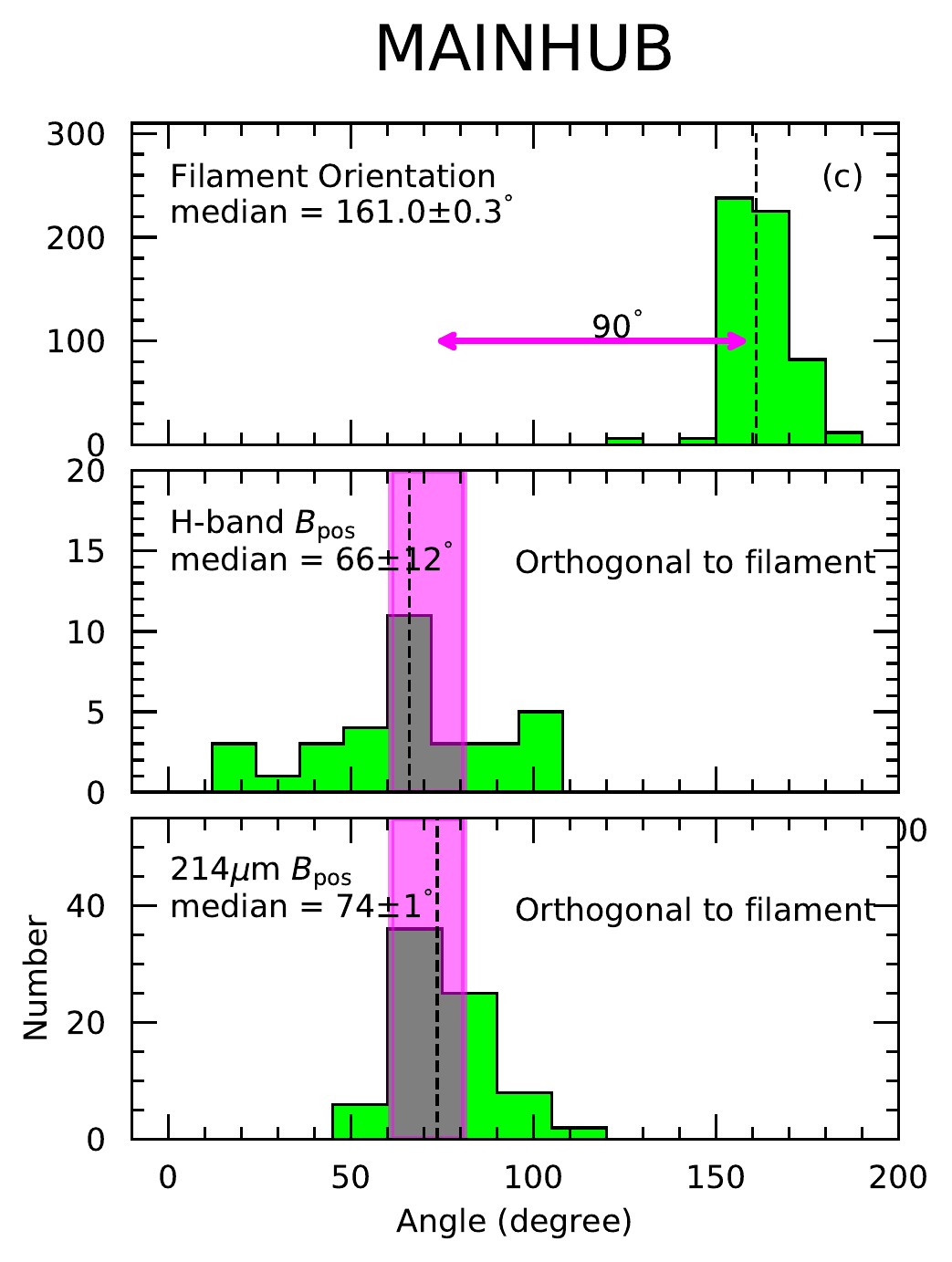}
\\
\includegraphics[width=0.5\linewidth,angle=0]{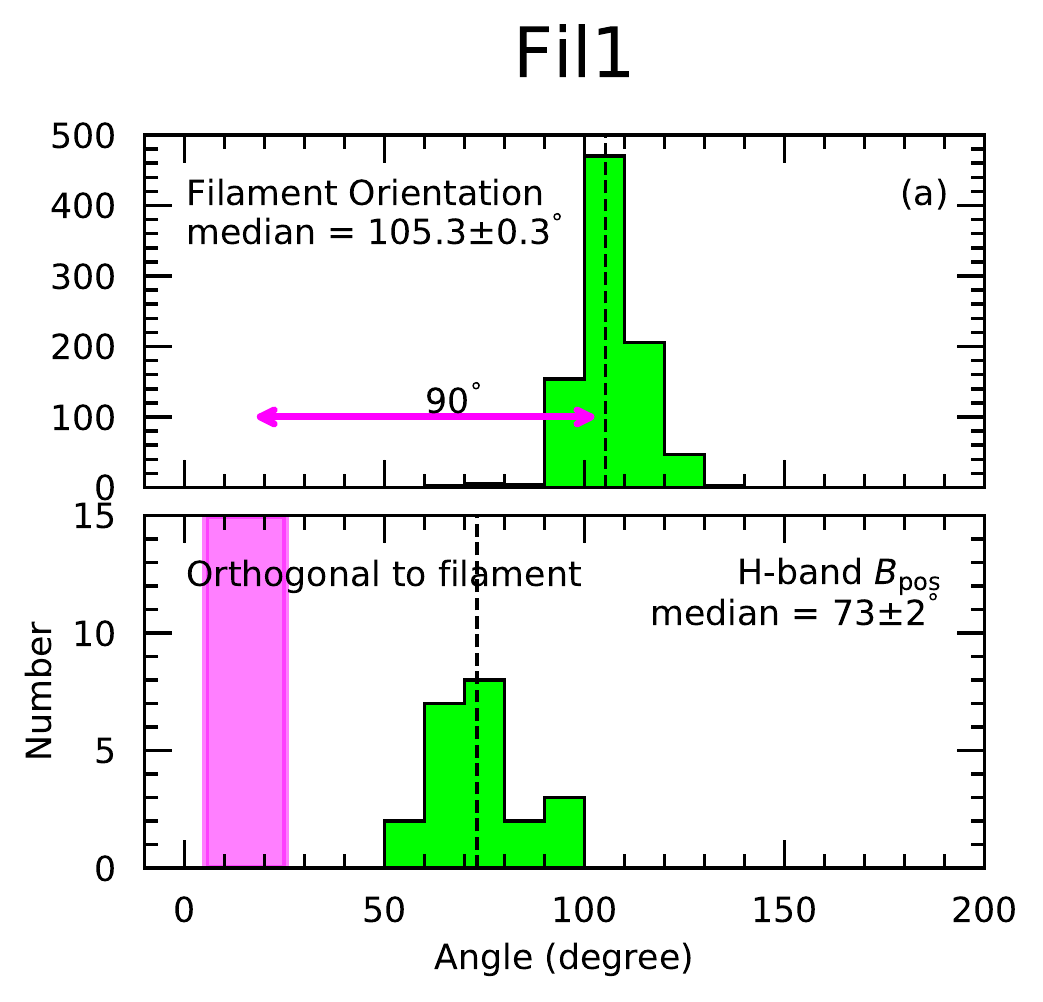}& \\
\end{tabular}
\centering
\caption{Distributions of filament angles derived from RHT
    analysis (top panels), magnetic field orientations in the NIR
    (second panels), and in the FIR (third panels, if present) for the FIL1 (a),
    FIL2 (b), and MAINHUB (c) elliptical regions. No HAWC+ detections have been found towards FIL1. The black dashed lines represent the medians of
    each distribution and the uncertainties shown are the errors of the means. The purple shaded bars represent the
    ranges that would conform to having either a parallel and
    perpendicular orientation of the field ($\pm 10$\degr) with respect
    to the filament angle.}
\label{fig:histo}
\end{figure}

We quantified the relation between cloud gas structures and magnetic fields
by measuring the relative projected orientations between the filaments and the
magnetic fields traced via polarimetry.
We characterized the filament orientations using the Rolling Hough
Transform (RHT\citep{clark2014:rht}), an image processing tool that determines orientations of linear
structures (Methods). 
We constructed histograms of RHT angles
and magnetic field orientations within the MAINHUB, FIL1, and FIL2 regions.
We also calculated median RHT angles and magnetic field orientations within each elliptical region. The results are shown in Fig.~\ref{fig:histo}. 
At the large spatial scales of the elliptical regions ($0.1$ -- $0.5$\,pc), 
the NIR polarization orientations for FIL1 show a narrow distribution whose 
median value is $32 \pm 2$\degr\ offset from being parallel to the RHT-traced gas filament
orientation. For FIL2, the median NIR field orientation is $73 \pm 5$\degr\ offset from the filament. For MAINHUB, the NIR field orientation is $95 \pm 12$\degr\ from the filament
angle. 

These offset angle values reveal the NIR-traced magnetic field is perpendicular to 
the gas structures within the FIL2 and MAINHUB zones. 
At smaller spatial scales ($<0.1$\,pc), however,  the smoothed HAWC+ 
observations of the FIL2 region show magnetic field orientations closer to being 
parallel to the gas filament elongation (medians offset by $22 \pm 3$\degr). 
For MAINHUB, the large-scale perpendicular relative orientation is preserved down to 
the smaller, FIR-traced scales (medians offset by $87 \pm 1$\degr).  
Different ellipse sizes were tried and found to yield consistent results.

FIL2, the FIR-brightest southern filament connecting to the central hub (see
Fig.~\ref{fig:overview}), thus shows a distinct change in magnetic field orientation, from being
perpendicular to the moderate column density gas structure to being parallel to the high column density one.  For every FIL2 NIR and FIR
dust polarization detection that met our SNR criteria, we extracted the 
corresponding magnetic field orientation (\Bpos) and the {\it Herschel}-based 
H$_2$ column density and converted the latter to visual extinction $A_V$. 
Fixing the filament orientation $X_{FIL}$ to the 144\degr\ median
value from Figure~\ref{fig:histo}, we 
computed its difference to each magnetic field orientation ($X_{\rm{}B_{pos}}-X_{FIL}$). The results are summarized in
Figure~\ref{fig:fil2_ang}. The Figure reveals the strong, systematic change in relative orientation between the magnetic field and the filament direction that
occurs near an extinction threshold $A_V\sim 21$~mag.

\section*{Discussion}

\paragraph{Evidence for magnetized filamentary accretion in FIL2}

At 10\,arcmin resolution, Planck Collaboration Int.\ XXXV\citep{Planck-Collaboration2016}
found, for Gould Belt molecular clouds, a change in relative orientations between 
magnetic fields and filament elongations, from being generally parallel to being 
perpendicular, occurs at \AV$>
2.7$\,mag. A similar analysis of the
Aquila rift\citep{soler2017}, which encompasses Serpens South, found an orientation 
transition at \AV$\sim 3.5$ mag. 
These studies revealed the dynamically important role played
by magnetic fields in collecting and channeling matter into present
molecular cloud configurations.  

\begin{figure*}[t]
  \centerline{\includegraphics[width=0.8\linewidth]{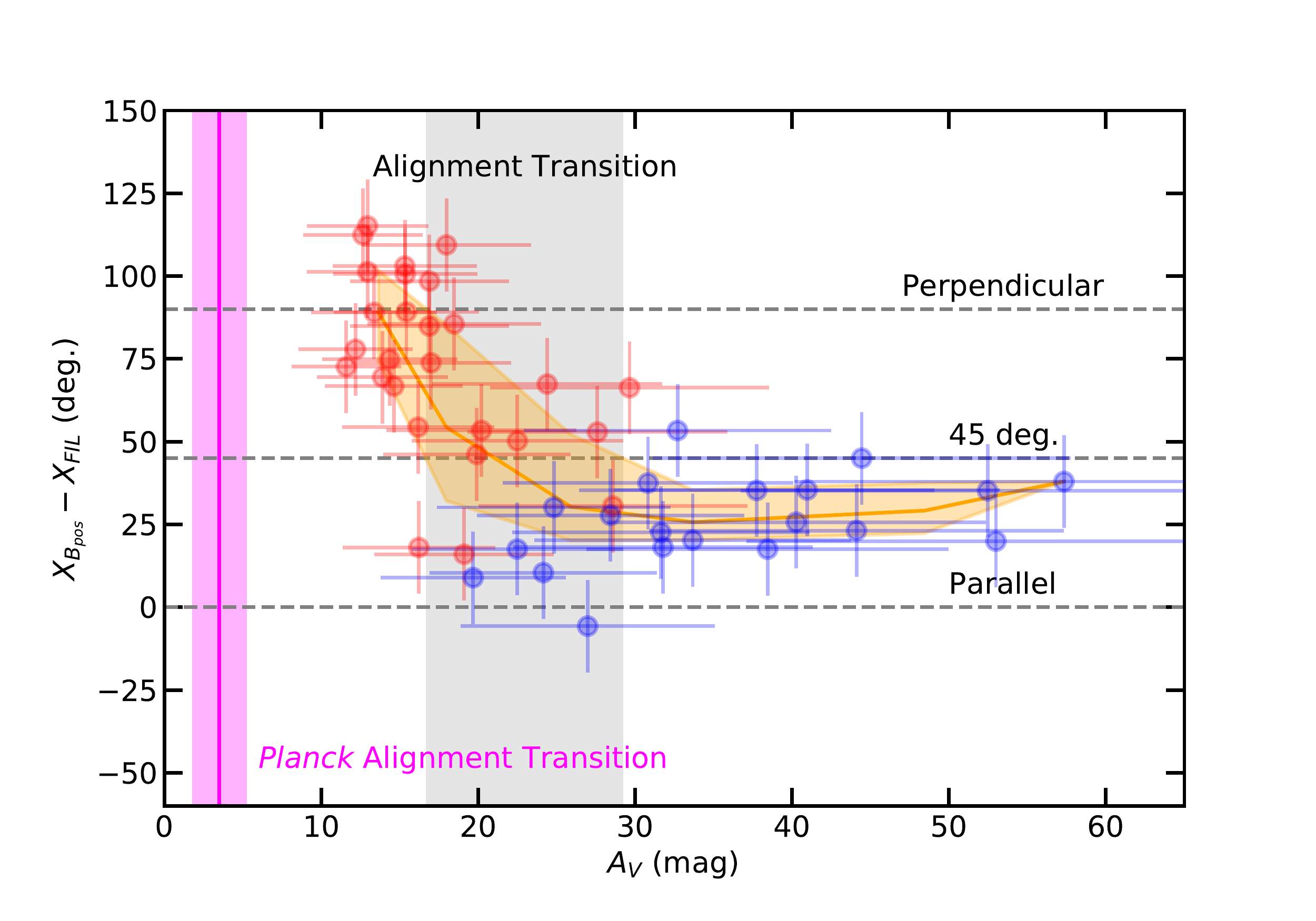}}
  \caption{Distribution of relative orientations of the FIL2 gas filament with
    respect to \Bpos\ as a function of $A_V$. The NIR \Bpos\ are shown as red filled circles and the FIR as blue
    filled circles.  Their corresponding 1\,$\sigma$ uncertainities are
    shown as errorbars. The three dashed lines show
    parallel, 45\degr, and perpendicular
    orientations of the magnetic field with the filament. The
    orange area captures the interquartile ranges for all data versus $A_V$.
The vertical gray rectangle
   highlights where the median relative orientation crosses 45\,\degr, with width
related to the uncertainty of the crossing value. This indicates where the
    transition from a perpendicular to parallel field
    orientation occurs. The magenta bar represents the lower $A_V$, first
    alignment transition suggested by {\it Planck} data, for the parallel to perpendicular transition near $A_V \sim 3.5$~mag \citep{soler2017}.}
\label{fig:fil2_ang}
\end{figure*}

The observations presented here of just such a star-forming filament
show that this picture is more complex. While the perpendicular magnetic field
orientation is preserved for sightlines piercing the dense hub, the moderate
column density FIL2 region filament exhibits a magnetic field 
orientation that undergoes a change back to being parallel to the filament.
The change in relative orientations between the magnetic field and the 
filament elongation occurs at about $A_V\sim 21$\,mag, as shown in Fig.~\ref{fig:fil2_ang}. 

This transition threshold was determined as follows.  
Adopted 30\%
  uncertainities for the pixels in the {\it Herschel}-based column density map, following
  Section 4.6 of Konyves et al.\citep{Konyves2015},
we binned the map pixels that matched to
NIR or FIR polarization detections
into logarithmically spaced bins of \AV. From the resulting distributions, we determined the
 first, second (median), and third quartiles for every bin. The span of the data
 that fell within the interquartile range, i.e. within the first and
 the third quartile, is shown as the orange shaded region in
 Fig.~\ref{fig:fil2_ang}. The transition threshold was taken to be
 where the median values of relative orientations cross 45\,\degr, namely 21~mag. 
The $A_V$ range of the transition is shown by the gray zone in the Figure. 
It is bounded by where the first quartiles of the low $A_V$
data points cross 45\degr, at 17~mag, and the where the third quartiles of 
the high $A_V$ data points similarly cross that line, at 29~mag.

We interpret this changing magnetic field orientation in the FIL2 region as evidence
for gravity dragging the denser gas, and entraining the frozen-in
large-scale magnetic field, to become a parallel flow of matter toward 
the MAINHUB region. In this interpretation, the FIL1, FIL2,
and FIL3 region filaments act as accretion channels to funnel
gas to the dense hub.  

Numerical simulations predict that magnetic field
orientations should follow such gravity-induced flows inside
dense filaments \citep{koertgen2015,gomez2018, li2018:mhd}. 
Magnetized accretion flows had been
expected and have been observed on protostellar envelope and smaller
disk size scales ($\le 500$\,AU)
\citep{sadavoy2018:alma2,maury2018,takahashi2019,legouellec2019}. 
Evidence for such accretion flows  have been found in the smooth velocity 
gradients in N$_2$H$^+$~1$\rightarrow$0  in the FIL2 region filament\citep{Kirk2013}(see also
Fernandez-Lopez et al.\citep{Fernandez-Lopez2014}).  Field-parallel orientations have
been observed recently in a dense filament in Orion's OMC1 region\citep{monsch2018} 
as well as in a distant, hub-filament 
infrared dark cloud\citep{liu2018}. 
The combined NIR and FIR polarimetric observations reported here show
that filamentary accretion
flows affect the local magnetic field orientation and thereby shape the magnetic field structure on filament size scales. 
A consistent picture emerges of a system of filaments merging into a
hub via gas flows along, or entraining, magnetic field lines.

\paragraph{Field Strengths}

Magnetic field support of filaments of the FIL2 region type must lose to 
gravitational collapse close to where the new change in
orientation has been discovered, in order to explain the onset of star formation. 
The mass-to-magnetic-flux ratio, $M/\Phi_B$, reaches its
critical value, $(M/\Phi_B)_{\rm{}cr}$, and induces a filament to
collapse and form stars, if $(M/\Phi_B)>(M/\Phi_B)_{\rm{}cr}$ (the magnetically 
supercritical condition). Following McKee and Ostriker\citep{mckee:araa07}, and recast into typical physical units
\citep[e.g.,][]{pillai2015}, 
\begin{equation}
\frac{(M/\Phi_B)}{(M/\Phi_B)_{\rm{}cr}} \approx 0.76\,\left(\frac{\langle N_{\rm
      H_{2}}\rangle}{10^{23}\,{\rm
      {cm}}^{-2}}\right)\left(\frac{B_{\rm{}tot}}{1~{}\rm{mG}}\right)^{-1}.\
\label{eq:mass-to-flux}
\end{equation}
For a transition at $A_V \sim 21$\,mag, 
$(M/\Phi_B)/(M/\Phi_B)_{\rm{}cr} \ge 1$ is fulfilled if $B_{\rm{}tot} \le
140$\,$\mu$G.  

We adopted a spheroidal density distribution and assumed flux--freezing\cite{mestel1966, myers2018} to obtain an estimate of the magnetic field strength for the MAINHUB region. These assumptions fully constrain the relative geometry of the density distribution and the magnetic field, but not their absolute scaling. We assumed that the dense core is magnetically supercritical within the $A_V\sim{}40$~mag contour (column density $\sim$4$\times{}10^{22}$\,cm$^{-2}$), where the central core becomes prominent relative to the surrounding material. This additional constrain fixes the distribution of density and magnetic field in absolute terms, resulting in a central magnetic field strength of $\sim{}870\,\rm{}\mu{}G$ and a central density of $\rho_0=6.3\times{}10^{5}~\rm{}cm^{-3}$ for the MAINHUB core in Serpens South.

This field strength estimate may be compared to values obtained from the empirical relation\citep{crutcher2012},  
$B=B_0\cdot{}(n_{\rm{}H_2}/10^4~{\rm{}cm^{-3}})^{0.65}$
with $B_0\lesssim{}150~\rm{}\mu{}G$. 
For the FIL2 region filament, we obtained a density 
estimate of $6.4\times10^4$\,\percc, appropriate for the transition 
at $A_V \sim 21$\,mag, through dividing the {\it Herschel} column density 
values there by an approximate filament width\citep{hill2012} of 
$\sim 0.1$~pc. Under these assumptions, the 
Crutcher relation suggests $B\lesssim 500$~$\mu$G, consistent with the 140~$\mu$G 
value obtained assuming magnetic supercriticality for the filament. 
For the MAINHUB region, the Crutcher 
relation suggests $B\lesssim 2$~mG at the modeled central 
density of $6.3\times{}10^{5}~\rm{}cm^{-3}$, while we estimated 
870~$\mu$G from flux-freezing.

That our estimates of $B$ are consistent 
with those suggested by the Crutcher relation is remarkable,
because our investigation probed (in particular in FIL2) gas that is less 
disturbed by star formation than the regions from which the Crutcher relation was derived (i.e., traced by masers and CN 
emission connected to very active star formation), and thus potentially 
more representative of the initial conditions for star formation.

\paragraph{Effects of protostellar radiation on fractional polarization}

While the mean magnetic field is well ordered along the NE-SW direction,
the FIR polarized fraction is not uniform in the MAINHUB region,
where the
greatest number of HAWC+ polarization detections are present. 
A significant decrease in
the polarization degree is evident in Fig.~\ref{fig:overview} towards
the brightest emission peak in MAINHUB. 

\begin{figure*}
\begin{tabular}{ll}
 \includegraphics[width=0.5\linewidth,angle=0]{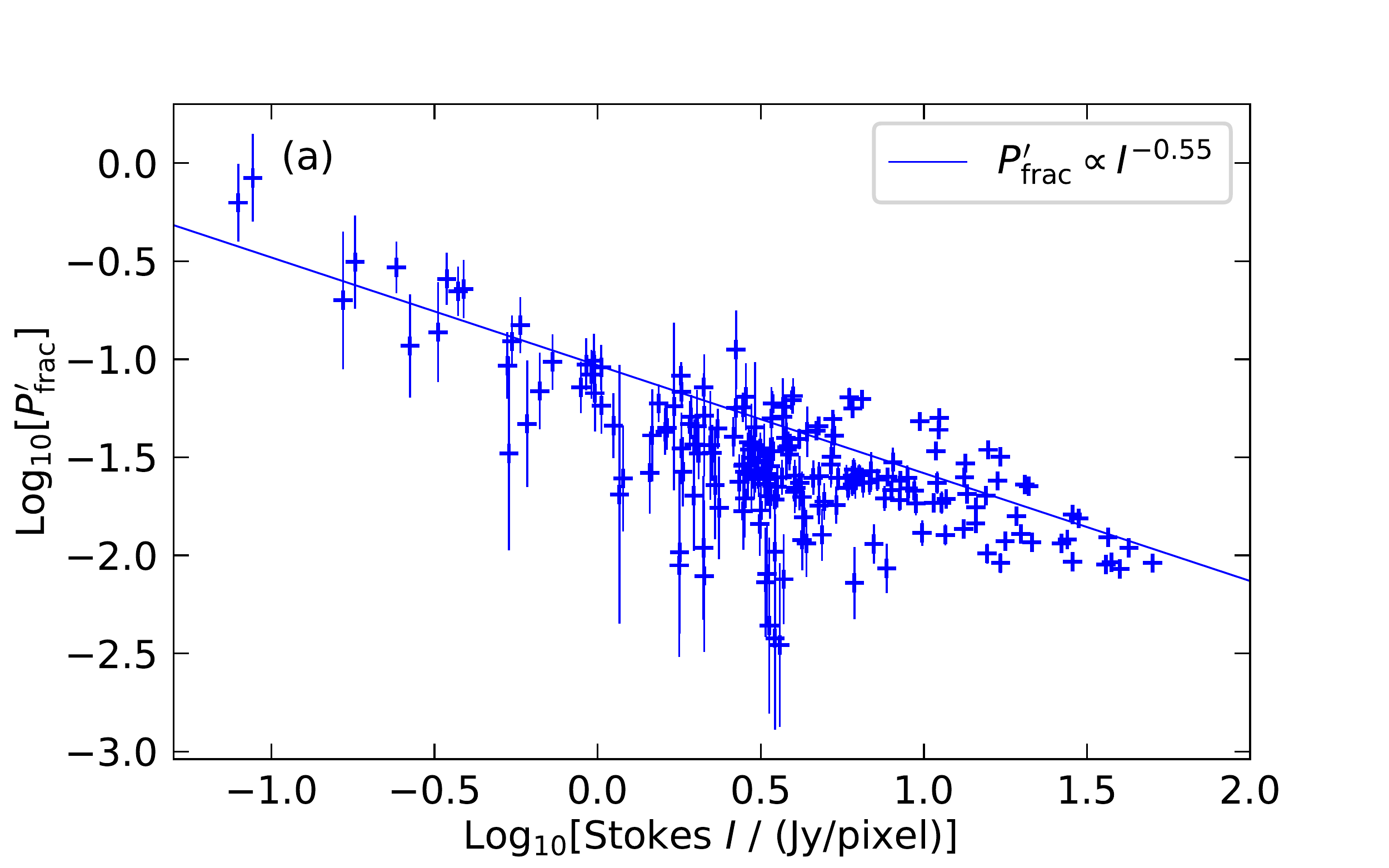}
  & \includegraphics[width=0.5\linewidth,angle=0]{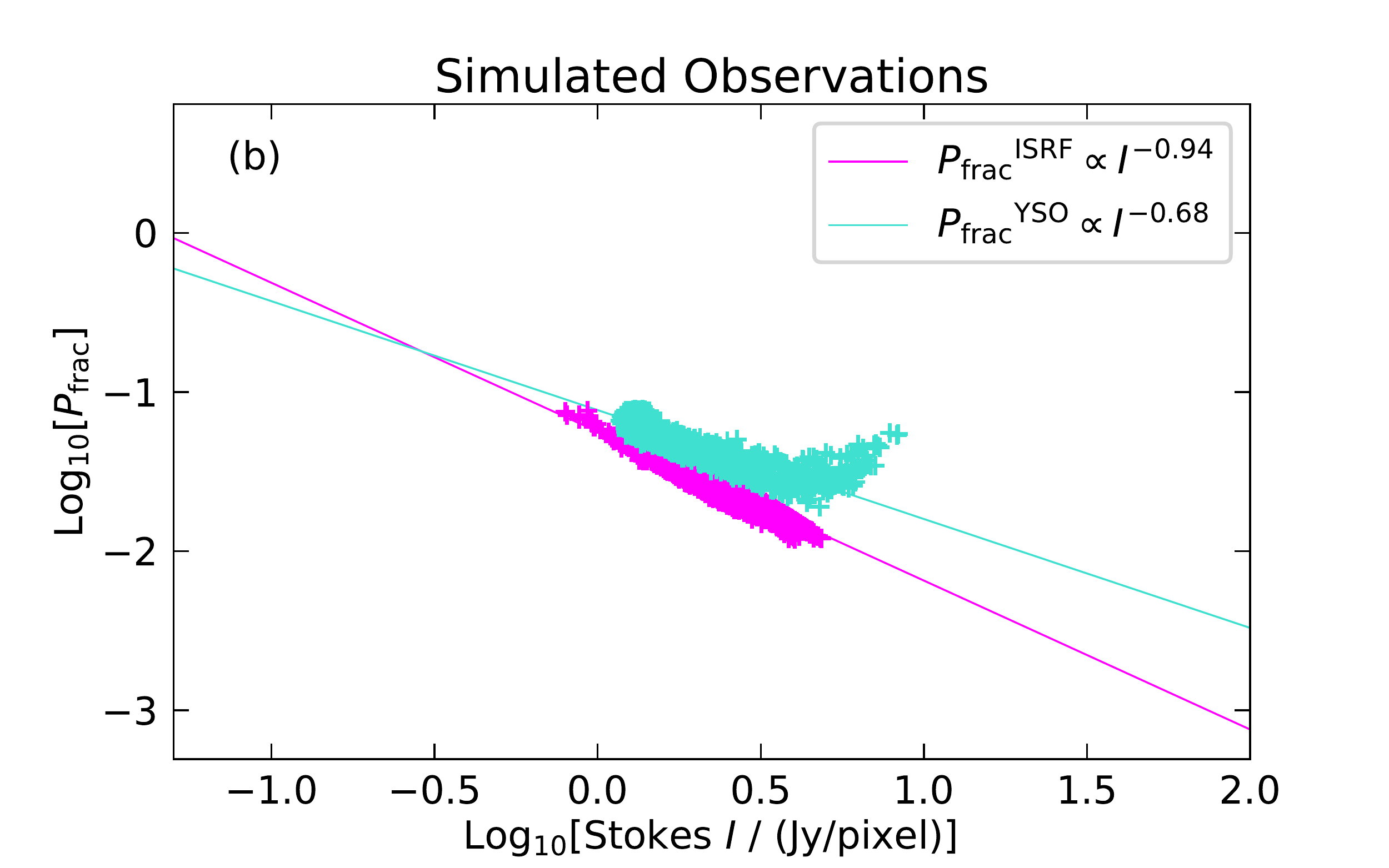}
  \\
\end{tabular}
\centering
\caption{Comparison of the correlation between fractional polarization and column density towards MAINHUB in Serpens South using HAWC+ 214\,$\mu$m (panel~[a]) and in simulated data (panel~[b]).   $P_{\rm frac}^{\prime}$ is the debiased
fractional polarization from observations and $P_{\rm frac}$ is the fractional polarization. Observational uncertainties are defined in the Methods section. The solid line in panel~(a) shows  a least square fit. Simulated observations from POLARIS in panel~(b) were conducted with ISRF alone (setup A, magenta crosses) and including the
    effect of a cluster of YSOs (setup B, turquoise crosses). The corresponding best fits to
    the model predictions are also shown.}
\label{fig:polaris}
\end{figure*}

In Figure~\ref{fig:polaris}, we show the debiased polarization fraction as a function of the FIR 214\,$\mu$m total intensity (Stokes $I$)
and the weighted least squares power-law fit to the data.  We do not apply any signal-to-noise cutoffs for
the fractional polarization. This is because polarization
fraction is a positive quantity, and follows a Rice distribution and therefore
would introduce a bias to high polarization values in
regions with low PSNR \citep{montier2015}. We derive a best-fit power-law slope  of 
$b = -0.55 \pm 0.03$. 

A relatively wide range in this slope has been
recently reported, with $b = -0.34$ towards the externally-illuminated Oph
A region \citep{pattle2019b:pol}, while starless cores such as Pipe
109/FeSt I-457  show\citep{alves2015:corr,kandori2018} $b \approx -1$.  
In well-illuminated
regions of molecular clouds, dust grains align with their
minor axes parallel to the local magnetic field direction. However,
because radiation does not sufficiently penetrate the denser parts of
clouds, the efficiency of the alignment with the local magnetic field may fall significantly (see review by Andersson et al.\citep{Andersson2015}). 
The observed slope of $b \sim -0.5$ in Serpens South is thus intermediate between the
two extreme cases: strong radiation inducing perfect dust grain alignment and weak radiation causing no alignment. The intermediate value for our slope  suggests that while the
efficiency of grain alignment decreases linearly with increasing
optical depth, FIR polarimetry still probes the magnetic field in the densest parts of this
cloud. This is a rather surprising result given the high extinction
towards the center of the Serpens South cluster, of the order $A_V \sim
150$\,mag. 

In order to probe the nature of the alignment in more detail, we
modeled the FIR dust polarization observations, based on the known properties
of the cloud's density profile, its temperature, and its star-formation state, 
as described in Methods. 
The modeling used the radiative transfer (RT) code POLARIS \citep[][]{reissl2016} 
to create simulated maps of dust temperature, RATs, and dust polarization for two setups. 
Setup ``A" used an external diffuse interstellar radiation field
(ISRF) as the only radiation source for dust
heating and grain alignment. Setup ``B" included the ISRF as well as 
internal sources of radiation from the protostars in the Serpens South Star Cluster (see Methods).

The resulting
correlations between polarization fraction $P_{\mathrm{frac}}$ and
Stokes $I$ are shown in Fig.\,\ref{fig:polaris}. 
Setup A shows a steeper slope because the external ISRF is extincted with increasing optical depth into the filament, causing the dust alignment
efficiency to decrease with $A_V$ or Stokes~$I$.  
In setup B, the radiation from the embedded YSOs increases the dust alignment in the 
cloud core center, leading to a shallower slope in the Figure.
The slope for our observations agrees better with the setup B model than with the 
setup A one. 
At the greater $A_V$ (Stokes~$I$) end of the Figure's right panel, 
the setup B distribution shows a rise in fractional polarization,
consistent with the localized influence of internal illumination on
grain alignment at the center of the cluster. However, this trend is
not resolved in our FIR data. 
In summary, embedded YSOs can cause a relatively shallow slope for the correlation, 
consistent with our FIR observations.

\section*{Conclusions}

Large-scale {\it Planck} observations of Gould Belt clouds find a transition
in relative orientation
between magnetic fields and gas filaments from being parallel for \AV$\le 2.7 - 3.5$\,mag to
being perpendicular beyond that \AV\ range.  Recent submm observations
show magnetic field orientations parallel to dense
filaments in a few star-forming regions \citep{monsch2018,liu2018},
but does not capture a transition. 

The Serpens South cloud is a dense, cluster-forming, hub-filament system in the Aquila
complex where we have used the HAWC+ FIR polarimeter instrument on SOFIA,
in conjunction with published NIR polarimetry, to discover a new transition 
occurring at even greater $A_V$ values. 
This new transition reveals the change of the magnetic field from 
being perpendicular to a large-scale, modest $A_V$, filament gas structure to becoming a smaller-scale magnetic field that is parallel to the densest gas structure 
beyond $A_V \sim$20~mag. In addition, the slope of the observed decrease in fractional
polarization with $A_V$ is better modeled when the external ISRF is augmented by
inclusion of illumination from the embedded protostars in the Serpens South cluster,
enhancing the link of magnetic fields traced by FIR polarimetry with the dense, opaque
gas.

This favors a scenario of gas filaments merging into a central hub, reorienting or 
entraining the magnetic field in the dense gas flows. 
Recent kinematic observations show evidence for gas flows
along such filaments towards the hub in systems, including
in Serpens South\citep{schneider2010:cygX, peretto2013,
  Kirk2013}. Thus, at $<0.1$\,pc scale, we find observational
evidence for the feeding of gas, containing significant magnetic fields, 
into a low-mass, cluster-forming region. 

\newpage
\section*{Acknowledgments}

Based in part on observations made with the NASA/DLR Stratospheric
Observatory for Infrared Astronomy (SOFIA). SOFIA is jointly operated
by the Universities Space Research Association, Inc. (USRA), under
NASA contract NAS2-97001, and the Deutsches SOFIA Institut (DSI) under
DLR contract 50 OK 0901 to the University of Stuttgart.
We sincerely thank the reviewers for their valuable comments that improved the clarity of this manuscript. 
Financial
support for this work was provided to TP by DLR through award \#~50~OR~1719.
Darren Dowell, the SOFIA HAWC+ team, the SOFIA  flight and ground crews, and the USRA SOFIA Project teams developed the SOFIA observatory, the HAWC+ instrument, performed
the airborne observations, processed and calibrated the data, and
delivered science-ready data products.
DPC acknowledges support funder NSF AST~18-14531, USRA SOF\textunderscore{}4-0026,
and NASA NNX15AE51G. DS acknowledges the support of the Bonn-Cologne Graduate School, which
is funded through the German Excellence Initiative and funding by the
Deutsche Forschungsgemeinschaft (DFG) via the Collaborative Research
Center SFB 956 “Conditions and Impact of Star Formation” (subprojects
C5 and C6). GAPF is partially supported by CNPq and FAPEMIG.

This research has made use of data from the Herschel Gould Belt survey
(HGBS) project (http://gouldbelt-herschel.cea.fr, Andr\'e et
al. 2010). 

\section*{Author contributions}
T.P. led the SOFIA proposal, data analysis, interpretation of the data, and the paper writing. Other authors contributed to
the writing of the manuscript. D.P.C., S.R., P.C.M., and
J.K. participated in the data analysis. E.L-R. led the SOFIA data
reduction. F.O.A. and G.A.F. conducted near-IR polarization
observations. K.S. provided the published near-IR polarization data used in this study. J.H., K.M.M, F.N, D.S., and H.W. provided expertise in molecular cloud studies.

\section*{Methods}

\paragraph{Observations} \label{sec:obs}

Serpens South was observed (PI: Pillai, ID: 05\_0206) on 2017 May 12-13 
using HAWC+ \citep{vaillancourt2007,dowell2010,harper2018} with
the 2.7-m SOFIA telescope. HAWC+ polarimetric observations
simultaneously imaged two orthogonal components of linear
polarization onto two detector arrays of up to $32 \times 40$ pixels
each. Observations performed using the $E$-band 
provided a pixel scale of 9.37\,\arcsec, a beam size of 18.2\,\arcsec\ and 
an instantaneous field-of-view of up to $4.2 \times 6.2$\,arcmin$^2$. 

Observations were conducted using the standard chop-nod polarimetric
mode. Specifically, a four-position dither pattern with an offset of
three pixels ($28\arcsec$) was used. At each dither position, a halfwave plate
stepped through four distinct position angles, and images were obtained at each step. 
Observations used nod times of 40\,s,
with a chop-frequency of 10.2\,Hz. Final polarimetric images were composed 
from observations of two different sky pointing directions. Based on the morphology of the source from \textit{Herschel} observations, chopping did not switch into significant flux contribution from diffuse extended emission or compact sources. 

The data were reduced using \textsc{hawc\_drp} pipeline v1.3.0 and
custom python routines. Detail data reduction steps can be found in Harper et al.
\citep{harper2018}. In summary, the raw data were demodulated,
chop-nod and background subtracted, flux calibrated, and Stokes
parameters were estimated along with their uncertainties. Fractional
polarization ($P$), position angle (PA),  and their uncertainties were evaluated using Eqs.~1, 2, 4, and 5 of Gordon et al.\cite{gordon2018}. The final $P$
and PA were corrected using measured instrumental polarizations and
were then debiased.  The debiased fractional polarization was derived
from the fractional polarization as\cite{serkowski1974}
$ P^{\prime}=\sqrt{(P^2-\sigma_{P}^{2})}$.

For regions of the cloud with weak detections (e.g., FIL2),
it was necessary to smooth the polarization data to increase
PSNR. The smoothing followed the approach detailed in Clemens et al.
\citep{clemens2018}. From a variety of trial
smoothing beams,
we chose a 
gaussian kernel $\rm{FWHM}=18.2$\,\arcsec  ($\sim$4 HAWC+ pixels) 
to smooth the SOFIA-delivered Stokes $QI$, $UI$, and $I$ maps. This yielded a
maximum
number of independent synthetic pixels with PSNR$ > 2.0$ ($\sigma_{PA} < 14^{\degr}$).

The near-IR $H$-band data were adopted from Sugitani et al.\citep{sugitani2011}.  
Only values that satisfied the following criteria were extracted: $P/\sigma_{P} > 3.0$ and
$P^*/3 < P < 3P^*$, where 
$P^* = 2.73([H - Ks] - 0.2)$ is the best-fit color dependence for $H$-band 
stars with $\sigma_{P} < 0.3$\,\%, following Kusune et al.\citep{kusune2019}.

\paragraph{Measuring Filament Orientations}

We quantified the relationships between cloud structures and magnetic fields
by measuring their relative projected orientations. To characterize filament orientations,
we used the RHT\citep{clark2014:rht}.

The regions containing the Serpens South filaments were delimited by 
ellipses. These ellipses were chosen to maximize the numbers of significant polarization detections contained within each while minimizing overlap between adjacent ellipses.  
Since the MAINHUB gas appears in relative isolation, there is little confusion with 
neighboring (FIL) ellipses. For it, we established
a large encompassing ellipse that extends well beyond the
high-column-density material in MAINHUB. 
This ensures that many NIR detections are contained in that zone. 
Narrower ellipses were chosen for FIL1
and FIL2 to avoid overlaps. Note that ellipse long-axis orientations are not 
meant to represent filament or hub gas structure elongations. The ellipses
merely define regions of interest associated with the dominant gas structures
contained within their boundaries.

To find linear structures in images and to determine their 
orientations, RHT uses three input parameters. Following the convention 
in Clark et al.\citep{clark2014:rht}, these are a smoothing kernel diameter (D$_{\rm K}$), a 
diameter of the window used to roll across the image (D$_{\rm W}$),  and an intensity 
threshold (Z) above which data are extracted.  Our input image was the {\it Herschel} dust 
column density map. Our choices of RHT parameters were 
$\rm{D}_K=55$~pixels ($= 5.7$\,arcmin),  
$\rm{D}_W=15$~pixels ($= 1.5$\,arcmin), and Z$=70\,\%$
of the map peak intensity.
We explored input variable values and found a wide combination of 
values still extracted the same filaments, ensuring robust results.

\paragraph{Flux-freezing model}

The Serpens South cloud harbors a centrally condensed core whose surrounding polarization directions suggest the ordered pattern expected for flux freezing 
 \citep{myers2018} combined with a random component which may be attributed to turbulent
motions. To estimate the maximum magnetic field strength associated with the star-forming Serpens South core, we assumed that the core is magnetically supercritical within the $A_V \sim 40$~mag contour 
(column density  $\sim 4\times{}10^{22}\rm{}cm^{-2}$), which is where the core appears to emerge from its harboring filament.  The field strength was obtained from the flux-freezing model \citep{mestel1966,myers2018}, and from a model of the core 
as prolate spheroid\citep{myers2017}.  

The core density model was obtained by fitting $p=2$ Plummer\citep{plummer1911} column
density profiles (consistent with the cloud structure\citep{Konyves2015}) to the principal axes of the {\it
  Herschel}-based core column density map shown in
Figure~\ref{fig:overview}. 
This modeling gave a central volume density
  $\rho_0=6.3\times{}10^{5}~\rm{}cm^{-3}$ and scale length
$R_{\rm flat} = 0.03~\rm{}pc$.
These parameters were used to obtain
  expressions for the core volume density and mean volume density, per Myers et al.\citep{myers2018}. 

Then, as shown in Myers et al.\citep{myers2018}, the peak field strength in the core equatorial plane can be expressed as
\begin{equation}
B_0 = \left[ (8 \pi /3) \, G^{1/2} \, \rho_0 \, r_0 \right]
\left[ 3 \, (\xi_c - \arctan \xi_c) \right]^{1/3} \, ,
\label{eq:myers18}
\end{equation}
where $\xi_c=x_c/r_0$ is the dimensionless radius of the $A_V \sim 40$~mag ($N_{H_2}=4\times{}10^{22}~\rm{}cm^{-2}$) map contour. For Serpens South, we found $\xi_c=2$. Evaluating equation (\ref{eq:myers18}) yielded $B_0=0.87\,\rm{}mG$,
which serves better than an order--of--magnitude estimate, based on the
assumption of the core being magnetically supercritical within the
$A_V \sim 40$~mag contour.  This estimate is useful, but we caution
that it harbors ill--constrained, difficult to quantify uncertainties.

\paragraph{Radiative Transfer using POLARIS}

We used the RT code POLARIS \cite[][]{reissl2016} to create simulated dust
temperature, RATs, and dust polarization maps on a cylindrical
grid. 
The cylindrical grid had a length of $1\ \mathrm{pc}$ and a maximum radius 
of $0.5\ \mathrm{pc}$. 
We applied a Plummer density profile 
${\rho(r)  = \rho_{0}[1+(r/R_{\mathrm{flat}})^2]^{-p/2}}$ 
with parameters
${R_{\mathrm{flat}}=0.03\ \mathrm{pc}}$ and $p=2$, 
from the previous section. 

The density $\rho_{0}$ was chosen, upon line of sight integration, to
match the observed central column density of $N_{\mathrm{H_2}}=1.6\times 10^{23}\ \mathrm{cm}^2$. 
This resulted in a total gas mass of $M_{\mathrm{gas}}=518\ \mathrm{M}_{\odot}$ within
the grid. 
The magnetic field was assumed to penetrate
perpendicular to the spine of the cylinder, with a constant strength of
$B=100\ \mu\mathrm{G}$. The gas temperature was a free parameter, 
set initially at $T_{\mathrm{gas}}=15\ \mathrm{K}$ to
encompass the upper limit of the measured gas temperature from
NH$_{3}$  observations \citep{friesen2016}.
 
For the dust properties, we used pre-calculated values of oblate grains with typical composition of $62.5\%$ graphite, $37.5\%$ silicate\citep{Reissl2017}, an aspect ratio of 1:2 and a mass ratio of $M_{\mathrm{dust}}/M_{\mathrm{gas}} = 0.01$. The grain size $a$ distribution followed ${n(a)\propto a^{-3.5}}$ with 
cut-offs at $a_{\mathrm{min}}=5\ \mathrm{nm}$ and $a_{\mathrm{max}}=500\ \mathrm{nm}$, typical for dense clouds \cite[e.g.,][]{martin1990}. 
 
We conducted two different RT simulation setups to calculate grain alignment efficiencies, following RAT theory \cite[see][]{reissl2016}.
 
For setup A, an external, diffuse ISRF \cite[][]{mathis1983} 
was the only radiation source. For setup B, the
luminosities and effective temperatures of the 37 Serpens South Class I YSOs
reported\citep{Dunham2015}, were included, in addition to the ISRF. 
The simulated
stars were placed randomly along the axial extent of the model cylinder, 
while their radial offsets from
the spine were drawn randomly from a Gaussian distribution
with a FWHM of $0.015\ \mathrm{pc}$. The latter was adopted
to be bounded by the stellar surface density distribution measured by
Gutermuth et al.\citep{gutermuth2008}.
Setup A resulted in a range of final dust temperatures, up to
$T_{\mathrm{dust}}=20\ \mathrm{K}$ at the edge of the filament model
and about $T_{\mathrm{dust}}=8\ \mathrm{K}$ toward the
spine. Regarding the RATs, dust grains larger than $a=181$~nm 
were well-aligned to the local magnetic field when close to the filament's surface, 
whereas even the
largest grains were not well-aligned at the location of the cylinder spine. 
The setup B YSOs brought RATs alignment to $a=125\ \mathrm{nm}$ in the close proximity of the YSOs.
Finally, we simulated synthetic polarization maps as being projected onto a
detector with $64\times 64$ pixels in a distance of $436\ \mathrm{pc}$
convolved with a beam of $19\ \mathrm{arcsec}$. 

\bibliographystyle{./naturemag}

\begin{thebibliography}{10}
\expandafter\ifx\csname url\endcsname\relax
  \def\url#1{\texttt{#1}}\fi
\expandafter\ifx\csname urlprefix\endcsname\relax\def\urlprefix{URL }\fi
\providecommand{\bibinfo}[2]{#2}
\providecommand{\eprint}[2][]{\url{#2}}

\bibitem{crutcher2012}
\bibinfo{author}{{Crutcher}, R.~M.}
\newblock \bibinfo{title}{{Magnetic Fields in Molecular Clouds}}.
\newblock \emph{\bibinfo{journal}{\araa}} \textbf{\bibinfo{volume}{50}},
  \bibinfo{pages}{29--63} (\bibinfo{year}{2012}).

\bibitem{li2014:ppvi_bfield}
\bibinfo{author}{{Li}, H.-B.} \emph{et~al.}
\newblock \bibinfo{title}{{The Link Between Magnetic Fields and Cloud/Star
  Formation}}.
\newblock \emph{\bibinfo{journal}{Protostars and Planets VI}}
  \bibinfo{pages}{101--123} (\bibinfo{year}{2014}).

\bibitem{Planck-Collaboration2016}
\bibinfo{author}{{Planck Collaboration XXXV}} \emph{et~al.}
\newblock \bibinfo{title}{{Planck intermediate results. XXXV. Probing the role
  of the magnetic field in the formation of structure in molecular clouds}}.
\newblock \emph{\bibinfo{journal}{\aap}} \textbf{\bibinfo{volume}{586}},
  \bibinfo{pages}{A138} (\bibinfo{year}{2016}).

\bibitem{soler2017}
\bibinfo{author}{{Soler}, J.~D.} \emph{et~al.}
\newblock \bibinfo{title}{{The relation between the column density structures
  and the magnetic field orientation in the Vela C molecular complex}}.
\newblock \emph{\bibinfo{journal}{\aap}} \textbf{\bibinfo{volume}{603}},
  \bibinfo{pages}{A64} (\bibinfo{year}{2017}).

\bibitem{sugitani2011}
\bibinfo{author}{{Sugitani}, K.} \emph{et~al.}
\newblock \bibinfo{title}{{Near-infrared-imaging Polarimetry Toward Serpens
  South: Revealing the Importance of the Magnetic Field}}.
\newblock \emph{\bibinfo{journal}{\apj}} \textbf{\bibinfo{volume}{734}},
  \bibinfo{pages}{63} (\bibinfo{year}{2011}).

\bibitem{Palmeirim2013}
\bibinfo{author}{{Palmeirim}, P.} \emph{et~al.}
\newblock \bibinfo{title}{{Herschel view of the Taurus B211/3 filament and
  striations: evidence of filamentary growth?}}
\newblock \emph{\bibinfo{journal}{\aap}} \textbf{\bibinfo{volume}{550}},
  \bibinfo{pages}{A38} (\bibinfo{year}{2013}).

\bibitem{Franco2015}
\bibinfo{author}{{Franco}, G.~A.~P.} \& \bibinfo{author}{{Alves}, F.~O.}
\newblock \bibinfo{title}{{Tracing the Magnetic Field Morphology of the Lupus I
  Molecular Cloud}}.
\newblock \emph{\bibinfo{journal}{\apj}} \textbf{\bibinfo{volume}{807}},
  \bibinfo{pages}{5} (\bibinfo{year}{2015}).

\bibitem{santos2016b}
\bibinfo{author}{{Santos}, F.~P.}, \bibinfo{author}{{Busquet}, G.},
  \bibinfo{author}{{Franco}, G. A.~P.}, \bibinfo{author}{{Girart}, J.~M.} \&
  \bibinfo{author}{{Zhang}, Q.}
\newblock \bibinfo{title}{{Magnetically Dominated Parallel Interstellar
  Filaments in the Infrared Dark Cloud G14.225-0.506}}.
\newblock \emph{\bibinfo{journal}{\apj}} \textbf{\bibinfo{volume}{832}},
  \bibinfo{pages}{186} (\bibinfo{year}{2016}).

\bibitem{soler2019}
\bibinfo{author}{{Soler}, J.~D.}
\newblock \bibinfo{title}{{Using Herschel and Planck observations to delineate
  the role of magnetic fields in molecular cloud structure}}.
\newblock \emph{\bibinfo{journal}{\aap}} \textbf{\bibinfo{volume}{629}},
  \bibinfo{pages}{A96} (\bibinfo{year}{2019}).

\bibitem{gutermuth2008}
\bibinfo{author}{{Gutermuth}, R.~A.} \emph{et~al.}
\newblock \bibinfo{title}{{The Spitzer Gould Belt Survey of Large Nearby
  Interstellar Clouds: Discovery of a Dense Embedded Cluster in the
  Serpens-Aquila Rift}}.
\newblock \emph{\bibinfo{journal}{\apjl}} \textbf{\bibinfo{volume}{673}},
  \bibinfo{pages}{L151--L154} (\bibinfo{year}{2008}).

\bibitem{ortiz-leon2018}
\bibinfo{author}{{Ortiz-Le{\'o}n}, G.~N.} \emph{et~al.}
\newblock \bibinfo{title}{{Gaia-DR2 Confirms VLBA Parallaxes in Ophiuchus,
  Serpens, and Aquila}}.
\newblock \emph{\bibinfo{journal}{\apjl}} \textbf{\bibinfo{volume}{869}},
  \bibinfo{pages}{L33} (\bibinfo{year}{2018}).

\bibitem{myers2009:fils}
\bibinfo{author}{{Myers}, P.~C.}
\newblock \bibinfo{title}{{Filamentary Structure of Star-forming Complexes}}.
\newblock \emph{\bibinfo{journal}{\apj}} \textbf{\bibinfo{volume}{700}},
  \bibinfo{pages}{1609--1625} (\bibinfo{year}{2009}).

\bibitem{lazarian2007}
\bibinfo{author}{{Lazarian}, A.} \& \bibinfo{author}{{Hoang}, T.}
\newblock \bibinfo{title}{{Radiative torques: analytical model and basic
  properties}}.
\newblock \emph{\bibinfo{journal}{\mnras}} \textbf{\bibinfo{volume}{378}},
  \bibinfo{pages}{910--946} (\bibinfo{year}{2007}).

\bibitem{Andersson2015}
\bibinfo{author}{{Andersson}, B.-G.}, \bibinfo{author}{{Lazarian}, A.} \&
  \bibinfo{author}{{Vaillancourt}, J.~E.}
\newblock \bibinfo{title}{{Interstellar Dust Grain Alignment}}.
\newblock \emph{\bibinfo{journal}{\araa}} \textbf{\bibinfo{volume}{53}},
  \bibinfo{pages}{501--539} (\bibinfo{year}{2015}).

\bibitem{kusune2019}
\bibinfo{author}{{Kusune}, T.} \emph{et~al.}
\newblock \bibinfo{title}{{Magnetic field structure in Serpens South}}.
\newblock \emph{\bibinfo{journal}{\pasj}}  (\bibinfo{year}{2019}).

\bibitem{andre2010:filaments}
\bibinfo{author}{{Andr{\'e}}, P.} \emph{et~al.}
\newblock \bibinfo{title}{{From filamentary clouds to prestellar cores to the
  stellar IMF: Initial highlights from the Herschel Gould Belt Survey}}.
\newblock \emph{\bibinfo{journal}{\aap}} \textbf{\bibinfo{volume}{518}},
  \bibinfo{pages}{L102+} (\bibinfo{year}{2010}).

\bibitem{Konyves2015}
\bibinfo{author}{{K{\"o}nyves}, V.} \emph{et~al.}
\newblock \bibinfo{title}{{A census of dense cores in the Aquila cloud complex:
  SPIRE/PACS observations from the Herschel Gould Belt survey}}.
\newblock \emph{\bibinfo{journal}{\aap}} \textbf{\bibinfo{volume}{584}},
  \bibinfo{pages}{A91} (\bibinfo{year}{2015}).

\bibitem{kauffmann2008}
\bibinfo{author}{{Kauffmann}, J.}, \bibinfo{author}{{Bertoldi}, F.},
  \bibinfo{author}{{Bourke}, T.~L.}, \bibinfo{author}{{Evans}, N.~J., II} \&
  \bibinfo{author}{{Lee}, C.~W.}
\newblock \bibinfo{title}{{MAMBO mapping of Spitzer c2d small clouds and
  cores}}.
\newblock \emph{\bibinfo{journal}{\aap}} \textbf{\bibinfo{volume}{487}},
  \bibinfo{pages}{993--1017} (\bibinfo{year}{2008}).

\bibitem{clark2014:rht}
\bibinfo{author}{{Clark}, S.~E.}, \bibinfo{author}{{Peek}, J.~E.~G.} \&
  \bibinfo{author}{{Putman}, M.~E.}
\newblock \bibinfo{title}{{Magnetically Aligned H I Fibers and the Rolling
  Hough Transform}}.
\newblock \emph{\bibinfo{journal}{\apj}} \textbf{\bibinfo{volume}{789}},
  \bibinfo{pages}{82} (\bibinfo{year}{2014}).

\bibitem{koertgen2015}
\bibinfo{author}{{K{\"o}rtgen}, B.} \& \bibinfo{author}{{Banerjee}, R.}
\newblock \bibinfo{title}{{Impact of magnetic fields on molecular cloud
  formation and evolution}}.
\newblock \emph{\bibinfo{journal}{\mnras}} \textbf{\bibinfo{volume}{451}},
  \bibinfo{pages}{3340--3353} (\bibinfo{year}{2015}).

\bibitem{gomez2018}
\bibinfo{author}{{G{\'o}mez}, G.~C.}, \bibinfo{author}{{V{\'a}zquez-Semadeni},
  E.} \& \bibinfo{author}{{Zamora-Avil{\'e}s}, M.}
\newblock \bibinfo{title}{{The magnetic field structure in molecular cloud
  filaments}}.
\newblock \emph{\bibinfo{journal}{\mnras}} \textbf{\bibinfo{volume}{480}},
  \bibinfo{pages}{2939--2944} (\bibinfo{year}{2018}).

\bibitem{li2018:mhd}
\bibinfo{author}{{Li}, P.~S.}, \bibinfo{author}{{Klein}, R.~I.} \&
  \bibinfo{author}{{McKee}, C.~F.}
\newblock \bibinfo{title}{{Formation of stellar clusters in magnetized,
  filamentary infrared dark clouds}}.
\newblock \emph{\bibinfo{journal}{\mnras}} \textbf{\bibinfo{volume}{473}},
  \bibinfo{pages}{4220--4241} (\bibinfo{year}{2018}).

\bibitem{sadavoy2018:alma2}
\bibinfo{author}{{Sadavoy}, S.~I.} \emph{et~al.}
\newblock \bibinfo{title}{{Dust Polarization toward Embedded Protostars in
  Ophiuchus with ALMA. II. IRAS 16293-2422}}.
\newblock \emph{\bibinfo{journal}{\apj}} \textbf{\bibinfo{volume}{869}},
  \bibinfo{pages}{115} (\bibinfo{year}{2018}).

\bibitem{maury2018}
\bibinfo{author}{{Maury}, A.~J.} \emph{et~al.}
\newblock \bibinfo{title}{{Magnetically regulated collapse in the B335
  protostar? I. ALMA observations of the polarized dust emission}}.
\newblock \emph{\bibinfo{journal}{\mnras}} \textbf{\bibinfo{volume}{477}},
  \bibinfo{pages}{2760--2765} (\bibinfo{year}{2018}).

\bibitem{takahashi2019}
\bibinfo{author}{{Takahashi}, S.} \emph{et~al.}
\newblock \bibinfo{title}{{ALMA High Angular Resolution Polarization Study: An
  Extremely Young Class 0 Source, OMC-3/MMS 6}}.
\newblock \emph{\bibinfo{journal}{\apj}} \textbf{\bibinfo{volume}{872}},
  \bibinfo{pages}{70} (\bibinfo{year}{2019}).

\bibitem{legouellec2019}
\bibinfo{author}{{Le Gouellec}, V. J.~M.} \emph{et~al.}
\newblock \bibinfo{title}{{Characterizing Magnetic Field Morphologies in Three
  Serpens Protostellar Cores with ALMA}}.
\newblock \emph{\bibinfo{journal}{\apj}} \textbf{\bibinfo{volume}{885}},
  \bibinfo{pages}{106} (\bibinfo{year}{2019}).

\bibitem{Kirk2013}
\bibinfo{author}{{Kirk}, H.} \emph{et~al.}
\newblock \bibinfo{title}{{Filamentary Accretion Flows in the Embedded Serpens
  South Protocluster}}.
\newblock \emph{\bibinfo{journal}{\apj}} \textbf{\bibinfo{volume}{766}},
  \bibinfo{pages}{115} (\bibinfo{year}{2013}).

\bibitem{Fernandez-Lopez2014}
\bibinfo{author}{{Fern{\'a}ndez-L{\'o}pez}, M.} \emph{et~al.}
\newblock \bibinfo{title}{{CARMA Large Area Star Formation Survey:
  Observational Analysis of Filaments in the Serpens South Molecular Cloud}}.
\newblock \emph{\bibinfo{journal}{\apjl}} \textbf{\bibinfo{volume}{790}},
  \bibinfo{pages}{L19} (\bibinfo{year}{2014}).

\bibitem{monsch2018}
\bibinfo{author}{{Monsch}, K.} \emph{et~al.}
\newblock \bibinfo{title}{{Dense Gas Kinematics and a Narrow Filament in the
  Orion A OMC1 Region Using NH$_{3}$}}.
\newblock \emph{\bibinfo{journal}{\apj}} \textbf{\bibinfo{volume}{861}},
  \bibinfo{pages}{77} (\bibinfo{year}{2018}).

\bibitem{liu2018}
\bibinfo{author}{{Liu}, T.} \emph{et~al.}
\newblock \bibinfo{title}{{A Holistic Perspective on the Dynamics of
  G035.39-00.33: The Interplay between Gas and Magnetic Fields}}.
\newblock \emph{\bibinfo{journal}{\apj}} \textbf{\bibinfo{volume}{859}},
  \bibinfo{pages}{151} (\bibinfo{year}{2018}).

\bibitem{mckee:araa07}
\bibinfo{author}{{McKee}, C.~F.} \& \bibinfo{author}{{Ostriker}, E.~C.}
\newblock \bibinfo{title}{{Theory of Star Formation}}.
\newblock \emph{\bibinfo{journal}{\araa}} \textbf{\bibinfo{volume}{45}},
  \bibinfo{pages}{565--687} (\bibinfo{year}{2007}).

\bibitem{pillai2015}
\bibinfo{author}{{Pillai}, T.} \emph{et~al.}
\newblock \bibinfo{title}{{Magnetic Fields in High-mass Infrared Dark Clouds}}.
\newblock \emph{\bibinfo{journal}{\apj}} \textbf{\bibinfo{volume}{799}},
  \bibinfo{pages}{74} (\bibinfo{year}{2015}).

\bibitem{mestel1966}
\bibinfo{author}{{Mestel}, L.}
\newblock \bibinfo{title}{{The magnetic field of a contracting gas cloud.
  I,Strict flux-freezing}}.
\newblock \emph{\bibinfo{journal}{\mnras}} \textbf{\bibinfo{volume}{133}},
  \bibinfo{pages}{265} (\bibinfo{year}{1966}).

\bibitem{myers2018}
\bibinfo{author}{{Myers}, P.~C.}, \bibinfo{author}{{Basu}, S.} \&
  \bibinfo{author}{{Auddy}, S.}
\newblock \bibinfo{title}{{Magnetic Field Structure in Spheroidal Star-forming
  Clouds}}.
\newblock \emph{\bibinfo{journal}{\apj}} \textbf{\bibinfo{volume}{868}},
  \bibinfo{pages}{51} (\bibinfo{year}{2018}).

\bibitem{hill2012}
\bibinfo{author}{{Hill}, T.} \emph{et~al.}
\newblock \bibinfo{title}{{Resolving the Vela C ridge with P-ArT{\'e}MiS and
  Herschel}}.
\newblock \emph{\bibinfo{journal}{\aap}} \textbf{\bibinfo{volume}{548}},
  \bibinfo{pages}{L6} (\bibinfo{year}{2012}).

\bibitem{montier2015}
\bibinfo{author}{{Montier}, L.} \emph{et~al.}
\newblock \bibinfo{title}{{Polarization measurement analysis. II. Best
  estimators of polarization fraction and angle}}.
\newblock \emph{\bibinfo{journal}{\aap}} \textbf{\bibinfo{volume}{574}},
  \bibinfo{pages}{A136} (\bibinfo{year}{2015}).

\bibitem{pattle2019b:pol}
\bibinfo{author}{{Pattle}, K.} \emph{et~al.}
\newblock \bibinfo{title}{{JCMT BISTRO Survey Observations of the Ophiuchus
  Molecular Cloud: Dust Grain Alignment Properties Inferred Using a Ricean
  Noise Model}}.
\newblock \emph{\bibinfo{journal}{\apj}} \textbf{\bibinfo{volume}{880}},
  \bibinfo{pages}{27} (\bibinfo{year}{2019}).

\bibitem{alves2015:corr}
\bibinfo{author}{{Alves}, F.~O.} \emph{et~al.}
\newblock \bibinfo{title}{{On the radiation driven alignment of dust grains:
  Detection of the polarization hole in a starless core (Corrigendum)}}.
\newblock \emph{\bibinfo{journal}{\aap}} \textbf{\bibinfo{volume}{574}},
  \bibinfo{pages}{C4} (\bibinfo{year}{2015}).

\bibitem{kandori2018}
\bibinfo{author}{{Kandori}, R.} \emph{et~al.}
\newblock \bibinfo{title}{{Distortion of Magnetic Fields in a Starless Core. V.
  Near-infrared and Submillimeter Polarization in FeSt 1-457}}.
\newblock \emph{\bibinfo{journal}{\apj}} \textbf{\bibinfo{volume}{868}},
  \bibinfo{pages}{94} (\bibinfo{year}{2018}).

\bibitem{reissl2016}
\bibinfo{author}{{Reissl}, S.}, \bibinfo{author}{{Wolf}, S.} \&
  \bibinfo{author}{{Brauer}, R.}
\newblock \bibinfo{title}{{Radiative transfer with POLARIS. I. Analysis of
  magnetic fields through synthetic dust continuum polarization measurements}}.
\newblock \emph{\bibinfo{journal}{\aap}} \textbf{\bibinfo{volume}{593}},
  \bibinfo{pages}{A87} (\bibinfo{year}{2016}).

\bibitem{schneider2010:cygX}
\bibinfo{author}{{Schneider}, N.} \emph{et~al.}
\newblock \bibinfo{title}{{Dynamic star formation in the massive DR21
  filament}}.
\newblock \emph{\bibinfo{journal}{\aap}} \textbf{\bibinfo{volume}{520}},
  \bibinfo{pages}{A49+} (\bibinfo{year}{2010}).

\bibitem{peretto2013}
\bibinfo{author}{{Peretto}, N.} \emph{et~al.}
\newblock \bibinfo{title}{{Global collapse of molecular clouds as a formation
  mechanism for the most massive stars}}.
\newblock \emph{\bibinfo{journal}{\aap}} \textbf{\bibinfo{volume}{555}},
  \bibinfo{pages}{A112} (\bibinfo{year}{2013}).

\bibitem{vaillancourt2007}
\bibinfo{author}{{Vaillancourt}, J.~E.} \emph{et~al.}
\newblock \bibinfo{title}{{Far-infrared polarimetry from the Stratospheric
  Observatory for Infrared Astronomy}} \textbf{\bibinfo{volume}{6678}},
  \bibinfo{pages}{66780D} (\bibinfo{year}{2007}).

\bibitem{dowell2010}
\bibinfo{author}{{Dowell}, C.~D.} \emph{et~al.}
\newblock \emph{\bibinfo{title}{{HAWCPol: a first-generation far-infrared
  polarimeter for SOFIA}}}, vol. \bibinfo{volume}{7735} of
  \emph{\bibinfo{series}{Society of Photo-Optical Instrumentation Engineers
  (SPIE) Conference Series}}, \bibinfo{pages}{77356H} (\bibinfo{year}{2010}).

\bibitem{harper2018}
\bibinfo{author}{{Harper}, D.~A.} \emph{et~al.}
\newblock \bibinfo{title}{{HAWC+, the Far-Infrared Camera and Polarimeter for
  SOFIA}}.
\newblock \emph{\bibinfo{journal}{Journal of Astronomical Instrumentation}}
  \textbf{\bibinfo{volume}{7}}, \bibinfo{pages}{1840008--1025}
  (\bibinfo{year}{2018}).

\bibitem{gordon2018}
\bibinfo{author}{{Gordon}, M.~S.} \emph{et~al.}
\newblock \bibinfo{title}{{SOFIA Community Science I: HAWC+ Polarimetry of 30
  Doradus}}.
\newblock \emph{\bibinfo{journal}{arXiv e-prints}}
  \bibinfo{pages}{arXiv:1811.03100} (\bibinfo{year}{2018}).

\bibitem{serkowski1974}
\bibinfo{author}{{Serkowski}, K.}
\newblock \emph{\bibinfo{title}{{Polarization techniques.}}}
  (\bibinfo{year}{1974}).

\bibitem{clemens2018}
\bibinfo{author}{{Clemens}, D.~P.} \emph{et~al.}
\newblock \bibinfo{title}{{Magnetic Field Uniformity Across the GF 9-2 YSO,
  L1082C Dense Core, and GF 9 Filamentary Dark Cloud}}.
\newblock \emph{\bibinfo{journal}{\apj}} \textbf{\bibinfo{volume}{867}},
  \bibinfo{pages}{79} (\bibinfo{year}{2018}).

\bibitem{myers2017}
\bibinfo{author}{{Myers}, P.~C.}
\newblock \bibinfo{title}{{Star-forming Filament Models}}.
\newblock \emph{\bibinfo{journal}{\apj}} \textbf{\bibinfo{volume}{838}},
  \bibinfo{pages}{10} (\bibinfo{year}{2017}).

\bibitem{plummer1911}
\bibinfo{author}{{Plummer}, H.~C.}
\newblock \bibinfo{title}{{On the problem of distribution in globular star
  clusters}}.
\newblock \emph{\bibinfo{journal}{\mnras}} \textbf{\bibinfo{volume}{71}},
  \bibinfo{pages}{460--470} (\bibinfo{year}{1911}).

\bibitem{friesen2016}
\bibinfo{author}{{Friesen}, R.~K.}, \bibinfo{author}{{Bourke}, T.~L.},
  \bibinfo{author}{{Di Francesco}, J.}, \bibinfo{author}{{Gutermuth}, R.} \&
  \bibinfo{author}{{Myers}, P.~C.}
\newblock \bibinfo{title}{{The Fragmentation and Stability of Hierarchical
  Structure in Serpens South}}.
\newblock \emph{\bibinfo{journal}{\apj}} \textbf{\bibinfo{volume}{833}},
  \bibinfo{pages}{204} (\bibinfo{year}{2016}).

\bibitem{Reissl2017}
\bibinfo{author}{{Reissl}, S.}, \bibinfo{author}{{Seifried}, D.},
  \bibinfo{author}{{Wolf}, S.}, \bibinfo{author}{{Banerjee}, R.} \&
  \bibinfo{author}{{Klessen}, R.~S.}
\newblock \bibinfo{title}{{The origin of dust polarization in molecular
  outflows}}.
\newblock \emph{\bibinfo{journal}{\aap}} \textbf{\bibinfo{volume}{603}},
  \bibinfo{pages}{A71} (\bibinfo{year}{2017}).

\bibitem{martin1990}
\bibinfo{author}{{Martin}, P.~G.} \& \bibinfo{author}{{Whittet}, D.~C.~B.}
\newblock \bibinfo{title}{{Interstellar Extinction and Polarization in the
  Infrared}}.
\newblock \emph{\bibinfo{journal}{\apj}} \textbf{\bibinfo{volume}{357}},
  \bibinfo{pages}{113} (\bibinfo{year}{1990}).

\bibitem{mathis1983}
\bibinfo{author}{{Mathis}, J.~S.}, \bibinfo{author}{{Mezger}, P.~G.} \&
  \bibinfo{author}{{Panagia}, N.}
\newblock \bibinfo{title}{{Interstellar radiation field and dust temperatures
  in the diffuse interstellar matter and in giant molecular clouds.}}
\newblock \emph{\bibinfo{journal}{\aap}} \textbf{\bibinfo{volume}{500}},
  \bibinfo{pages}{259--276} (\bibinfo{year}{1983}).

\bibitem{Dunham2015}
\bibinfo{author}{{Dunham}, M.~M.} \emph{et~al.}
\newblock \bibinfo{title}{{Young Stellar Objects in the Gould Belt}}.
\newblock \emph{\bibinfo{journal}{\apjs}} \textbf{\bibinfo{volume}{220}},
  \bibinfo{pages}{11} (\bibinfo{year}{2015}).

\end{thebibliography}

\end{document}